\documentclass{iopjournal}

\begin{document}


\title{A Symplectic Map Approach to Magnetic Field-Line Dynamics in Tokamaks}

\author{Diego F.M. Oliveira$^{*1}$\orcid{0000-0002-2375-0539}, Edson D.\ Leonel$^2$\orcid{0000-0001-8224-3329} }

\affil{$^1$School of Electrical Engineering and Computer Science, University of North Dakota, Grand Forks, ND, USA. }

\affil{$^2$Departamento de F\'isica, Unesp - Universidade Estadual Paulista -  Av.24A. 1515, 13506-700, Rio Claro, SP, Brazil}

\affil{$^*$Author to whom any correspondence should be addressed.}

\email{diegofregolente@gmail.com}


\begin{abstract}
Magnetic field-line transport in tokamaks is governed by the interplay between chaotic dynamics and invariant phase-space structures that act as partial barriers to radial motion. Understanding how these structures influence transport over long time scales is essential for describing magnetic confinement in toroidal plasmas. In this work, we investigate the conservative Tokamap, an exact symplectic mapping for magnetic field-line dynamics, using a unified geometrical, dynamical, and statistical framework. Phase-space portraits reveal the coexistence of invariant tori, magnetic island chains, and chaotic regions characteristic of mixed Hamiltonian systems, while the largest Lyapunov exponent provides a quantitative measure of local dynamical instability. Ensemble-averaged transport exhibits dynamic scaling characterized by growth, saturation, and crossover regimes connected through a generalized homogeneous scaling theory. The resulting scaling exponents satisfy the predicted scaling relation and produce an excellent collapse of the transport curves onto a universal function. To characterize asymptotic transport, we analyze Poincaré recurrence statistics and show that decreasing the magnetic-shear parameter systematically shifts the survival probability toward longer recurrence times, indicating progressively slower transport. The characteristic transport time follows the algebraic scaling $\tau_c\propto x_q^{-0.213}$, demonstrating that long-time dynamics are increasingly dominated by stickiness associated with KAM islands, resonance chains, and cantori. These results establish a direct connection between the geometrical organization of Hamiltonian phase space and macroscopic transport properties, showing that transport efficiency is controlled not only by local chaotic instability but also by the global structure of invariant transport barriers. The present framework provides a comprehensive description of magnetic field-line transport in the conservative Tokamap and offers a general approach for investigating long-time transport in Hamiltonian systems with mixed phase space.
\end{abstract}

\section{Introduction}

Magnetic confinement fusion seeks to sustain thermonuclear reactions by restricting the motion of high-temperature plasmas through externally generated magnetic fields. Among the available confinement concepts, the tokamak remains the most extensively investigated owing to its ability to achieve long confinement times and high plasma temperatures \cite{wesson2011,ongena2016magnetic,stacey2010fusion,miyamoto2005plasma}. In an ideal axisymmetric equilibrium, charged particles execute a rapid gyromotion around magnetic field lines while their guiding centers drift along nested toroidal magnetic surfaces. Because transport across magnetic field lines is much slower than transport along them, these nested magnetic surfaces act as effective barriers to radial particle and energy transport and therefore constitute the foundation of magnetic confinement~\cite{wesson2011,white2014,hazeltine2003,freidberg2014}.

In realistic devices, however, perfectly nested magnetic surfaces are never achieved. Error fields, resonant magnetic perturbations, tearing modes, kink instabilities, edge-localized modes, and externally applied control fields modify the magnetic topology \cite{freidberg2008plasma}. Near rational magnetic surfaces satisfying $q=m/n$, these perturbations generate chains of magnetic islands that broaden and eventually overlap as the perturbation strength increases, producing stochastic magnetic layers in which field lines undergo irregular radial excursions. The resulting degradation of magnetic surfaces enhances radial transport, alters heat deposition patterns, weakens transport barriers, and ultimately limits plasma confinement~\cite{rechester1978,white2014,abdullaev2006,evans2006edge,evans2006physics,loarte2007chapter}.

The transition from ordered magnetic surfaces to stochastic magnetic fields is one of the central problems in magnetic confinement physics and, simultaneously, a paradigmatic example of Hamiltonian chaos. Its theoretical foundations are provided by the Kolmogorov--Arnold--Moser (KAM) theory, resonance-overlap arguments, and the theory of transport in Hamiltonian systems with mixed phase space~\cite{kolmogorov1954,arnold1963,moser1962,chirikov1979,lichtenberg1992,meiss1992,leonel2016thermodynamics}. KAM theory predicts that sufficiently irrational invariant tori survive weak perturbations and continue to impede radial transport, whereas stronger perturbations progressively destroy these barriers. Chirikov's resonance-overlap criterion provides a practical estimate for the onset of large-scale stochasticity~\cite{chirikov1979}, while Greene's residue criterion predicts the breakup of selected invariant tori~\cite{greene1979}. Even after breakup, remnants of invariant tori survive as cantori, producing stickiness, anomalous diffusion, and long recurrence times by acting as partial transport barriers~\cite{mackay1984,meiss1992,zaslavsky2002,zaslavsky2007,altmann2005stickiness}.

Magnetic field-line trajectories naturally admit a Hamiltonian formulation whenever one magnetic coordinate is adopted as the independent variable. In toroidal geometry, the toroidal angle plays the role of time, while the poloidal angle and a magnetic-flux coordinate form a canonical pair, yielding a non-autonomous Hamiltonian system with one-and-one-half degrees of freedom~\cite{boozer1983,cary1983,white2014,abdullaev2006}. This formulation follows directly from the divergence-free nature of the magnetic field and establishes a direct connection between magnetic confinement and Hamiltonian dynamics. Consequently, magnetic field-line trajectories preserve phase-space area and can be efficiently represented through symplectic Poincar\'e maps, which replace continuous integration by successive intersections with a fixed poloidal section while preserving the underlying Hamiltonian structure. Such maps have become indispensable tools for investigating magnetic islands, stochastic layers, transport barriers, chaotic diffusion, stickiness, and recurrence statistics in magnetically confined plasmas~\cite{abdullaev2006,constantinescu2005non,hudson2012computation}.

Over the past several decades, numerous symplectic mappings have been developed to model magnetic field-line transport in toroidal plasmas \cite{abdullaev2006construction,ullmann2000symplectic}. Early mappings demonstrated that relatively simple canonical transformations reproduce the principal nonlinear mechanisms governing magnetic confinement while remaining computationally inexpensive~\cite{morrison2000magnetic,abdullaev2006,white2014,abdullaev2014magnetic}. These reduced descriptions have been successfully applied to the study of ergodic magnetic limiters, divertor configurations, stochastic edge layers, wall connection lengths, magnetic footprints, and transport in perturbed toroidal geometries.

Among these reduced models, the Tokamap introduced by Balescu, Vlad, and Spineanu has become one of the most influential descriptions of magnetic field-line dynamics in tokamaks~\cite{balescu1998tokamap}. Derived directly from Hamiltonian principles, the Tokamap preserves exact symplecticity while capturing the essential geometrical features of toroidal magnetic confinement. Despite its mathematical simplicity, it reproduces many of the nonlinear phenomena observed in realistic magnetic configurations, including magnetic island formation, resonance overlap, stochastic transport, partial transport barriers, stickiness, and the progressive destruction of invariant magnetic surfaces. Its combination of physical realism, computational efficiency, and analytical tractability has established the Tokamap as one of the standard models for investigating nonlinear transport in magnetically confined plasmas.

The original Tokamap has subsequently been extended in several directions to improve both its numerical accuracy and its ability to describe more realistic magnetic equilibria \cite{eberhard2005symmetric,bartoloni2016shearless}. Symmetric Tokamaps were introduced to provide a more faithful representation of the underlying Hamiltonian flow, bounded Tokamaps ensure physically admissible radial motion, and generalized versions incorporating non-monotonic safety-factor profiles have enabled the investigation of internal transport barriers and reversed magnetic shear~\cite{balescu2003,abdullaev2006,wingen2005stochastic}. These developments have considerably broadened the applicability of Tokamap models and have provided valuable insight into anomalous diffusion, subdiffusive transport, transport barriers, escape dynamics, intermittency, and the role of invariant manifolds in magnetic confinement~\cite{misguich2001dynamics}.

A particularly important extension is the reversed-shear Tokamap (revtokamap), introduced to describe magnetic configurations with non-monotonic safety-factor profiles~\cite{balescu1998revtokamap}. Unlike conventional twist maps, the revtokamap violates the twist condition because the rotation number possesses an extremum. This modification gives rise to phenomena absent in ordinary Hamiltonian twist systems, including separatrix reconnection, island collisions, shearless invariant curves, and exceptionally robust internal transport barriers~\cite{shinohara1998indicators}. These studies further established Tokamap-based models as a natural bridge between magnetic confinement physics and the broader theory of Hamiltonian transport in twist and nontwist dynamical systems.

Despite the extensive literature on Tokamap models, several aspects of long-time Hamiltonian transport remain incompletely understood. While Poincar\'e sections reveal the coexistence of invariant tori, resonance islands, and chaotic seas, they provide only a qualitative description of the dynamics. A more complete characterization requires complementary dynamical and statistical tools capable of quantifying local instability and long-time transport. Lyapunov exponents provide a direct measure of the exponential divergence of nearby trajectories and therefore identify chaotic regions of phase space, whereas recurrence statistics quantify the influence of sticky trajectories associated with KAM islands, cantori, and partial transport barriers. Together with scaling analyses, these quantities provide a comprehensive description of transport in Hamiltonian systems with mixed phase space.

In the present work, we investigate the conservative Tokamap through a unified geometrical, dynamical, and statistical approach. We first characterize the global phase-space organization using Poincar\'e sections and identify the principal regular and chaotic structures. We then compute Lyapunov exponents to quantify the evolution of dynamical instability over a broad range of perturbation strengths and investigate scaling properties associated with radial transport. Finally, we analyze Poincar\'e recurrence statistics to characterize long-time transport and the influence of stickiness on chaotic trajectories. Together, these complementary analyses provide a comprehensive characterization of the relationship between phase-space geometry, local instability, and transport in one of the most widely used Hamiltonian models of magnetic field-line dynamics.

The dynamics is formulated in terms of the action--angle variables $(\Psi,T)$, where $\Psi$ denotes the normalized toroidal magnetic flux and $T$ the normalized poloidal angle. The radial coordinate satisfies $0\leq\Psi\leq1$, with $\Psi=0$ corresponding approximately to the magnetic axis and $\Psi=1$ to the plasma boundary, while $T\in[0,1)$ represents the normalized poloidal angle. In the following section, we derive the conservative Tokamap from its Hamiltonian formulation and subsequently investigate its phase-space organization, Lyapunov stability, scaling properties, and long-time transport through Poincar\'e recurrence statistics.

\section{The Conservative Tokamap}

Magnetic field-line dynamics in toroidal plasmas can be efficiently investigated using symplectic mappings, which provide a computationally inexpensive alternative to the direct integration of the continuous field-line equations. Rather than following magnetic field lines continuously, these mappings describe the successive intersections of a field line with a fixed poloidal cross section through an iterative canonical transformation. Since they preserve the Hamiltonian structure of the underlying dynamics, symplectic maps accurately reproduce the long-time evolution of magnetic field lines while requiring only a fraction of the computational cost of direct numerical integration. Consequently, they have become one of the principal theoretical tools for investigating magnetic islands, resonance overlap, stochastic transport, transport barriers, chaotic diffusion, and long-time magnetic confinement in toroidal plasmas~\cite{abdullaev2006,white2014}.

Among the various symplectic mappings proposed for magnetic confinement studies, the Tokamap introduced by Balescu \emph{et al.}~\cite{balescu1998tokamap} has become one of the most successful reduced models of magnetic field-line dynamics. Derived directly from the Hamiltonian formulation of magnetic field-line equations, the Tokamap preserves the symplectic character of the continuous system while retaining the essential nonlinear mechanisms governing magnetic transport. Despite its relatively simple mathematical structure, it reproduces magnetic island formation, resonance overlap, stochastic transport, partial transport barriers, and the progressive destruction of invariant magnetic surfaces, making it an ideal framework for investigating nonlinear transport in tokamak plasmas.

The dynamics are formulated in terms of the canonical action--angle variables $(\Psi,T)$, where $\Psi$ denotes the normalized toroidal magnetic flux and $T$ is the normalized poloidal angle. The radial coordinate satisfies $0\leq\Psi\leq1$, where $\Psi=0$ corresponds approximately to the magnetic axis and $\Psi=1$ to the plasma boundary. The angular coordinate is periodic, $
0\leq T<1$,so that $T$ is defined modulo unity.

The magnetic field-line dynamics are generated by the Hamiltonian

\begin{equation}
H(\Psi,T)=H_0(\Psi)+H_1(\Psi,T),
\label{eq:Hamiltonian}
\end{equation}
where the integrable contribution $H_0(\Psi)$ describes the equilibrium magnetic configuration, while the perturbation $H_1(\Psi,T)$ represents the magnetic perturbation responsible for the formation of resonant magnetic islands and stochastic transport.

In the absence of perturbations, magnetic field lines remain confined to invariant magnetic surfaces and satisfy

\begin{equation}
\frac{dT}{d\phi}
=
\frac{1}{q(\Psi)},
\label{eq:qprofile}
\end{equation}
where $\phi$ denotes the toroidal angle and $q(\Psi)$ is the safety-factor profile. Physically, the safety factor measures the number of toroidal revolutions performed by a magnetic field line during one complete poloidal turn and therefore determines the local winding of magnetic field lines around the torus. Resonances occur whenever

\begin{equation}
q(\Psi)=\frac{m}{n},
\end{equation}
where $m$ and $n$ are integers. At these rational magnetic surfaces, perturbations generate magnetic island chains that dominate the nonlinear dynamics of the system.

Following the original construction of Balescu \emph{et al.}~\cite{balescu1998tokamap}, the perturbation Hamiltonian is chosen as

\begin{equation}
H_1(\Psi,T)
=
-
\frac{x_L}{4\pi^2}
\frac{\Psi}{1+\Psi}
\cos(2\pi T),
\label{eq:H1}
\end{equation}
where the dimensionless parameter $x_L$ controls the strength of the magnetic perturbation. As $x_L$ increases, magnetic islands broaden, neighboring resonances begin to overlap, and the dynamics evolve from predominantly regular motion toward global stochastic transport.

The evolution of magnetic field lines is obtained from Hamilton's canonical equations,

\begin{equation}
\frac{d\Psi}{d\phi}
=
-
\frac{\partial H}{\partial T},
\qquad
\frac{dT}{d\phi}
=
\frac{\partial H}{\partial\Psi},
\label{eq:HamiltonEq}
\end{equation}
which preserve phase-space area as a direct consequence of the Hamiltonian nature of the system. This symplectic property is fundamental because it guarantees that the discrete mapping faithfully reproduces the geometrical structure of the continuous dynamics without introducing artificial numerical dissipation, making the Tokamap particularly suitable for investigating long-time magnetic field-line transport.

Integrating Eq.~(\ref{eq:HamiltonEq}) over one complete toroidal revolution generates a canonical transformation relating two successive intersections of a magnetic field line with the chosen poloidal section. Following the construction introduced by Balescu \emph{et al.}~\cite{balescu1998tokamap}, the mapping is obtained by discretizing the Hamiltonian equations while preserving their symplectic structure.

Since the equilibrium Hamiltonian satisfies

\begin{equation}
\frac{dH_0}{d\Psi}
=
\frac{1}{q(\Psi)},
\label{eq:H0}
\end{equation}
the complete Hamiltonian becomes

\begin{equation}
H(\Psi,T)
=
\int^\Psi
\frac{d\Psi'}{q(\Psi')}
-
\frac{x_L}{4\pi^2}
\frac{\Psi}{1+\Psi}
\cos(2\pi T),
\label{eq:Hcomplete}
\end{equation}
where the first term describes the unperturbed rotation of magnetic field lines and the second term introduces the resonant magnetic perturbation.

Substituting Eq.~(\ref{eq:Hcomplete}) into Hamilton's equations yields

\begin{equation}
\frac{d\Psi}{d\phi}
=
-
\frac{x_L}{2\pi}
\frac{\Psi}{1+\Psi}
\sin(2\pi T),
\label{eq:dpsi}
\end{equation}
and
\begin{equation}
\frac{dT}{d\phi}
=
\frac{1}{q(\Psi)}
-
\frac{x_L}{4\pi^2}
\frac{\cos(2\pi T)}
{(1+\Psi)^2}.
\label{eq:dtheta}
\end{equation}

The radial equation is first discretized over one toroidal period. To preserve the symplectic character of the mapping, the nonlinear term is evaluated at the updated radial coordinate, leading to the implicit relation

\begin{equation}
\Psi_{n+1}
=
\Psi_n
-
\frac{x_L}{2\pi}
\frac{\Psi_{n+1}}
{1+\Psi_{n+1}}
\sin(2\pi T_n).
\label{eq:implicit}
\end{equation}

Because the unknown variable $\Psi_{n+1}$ appears on both sides of Eq.~(\ref{eq:implicit}), the equation cannot be solved directly. Multiplying both sides by $(1+\Psi_{n+1})$ transforms the expression into a quadratic polynomial in $\Psi_{n+1}$. Introducing the auxiliary quantity

\begin{equation}
P_n
=
\Psi_n
-
1
-
\frac{x_L}{2\pi}
\sin(2\pi T_n),
\label{eq:P}
\end{equation}
the quadratic equation assumes the compact form

\begin{equation}
\Psi_{n+1}^2
-
P_n\Psi_{n+1}
-
\Psi_n
=
0.
\label{eq:quadratic}
\end{equation}

Application of the quadratic formula yields two mathematical solutions. Since the magnetic flux must remain non-negative, only the positive branch is physically admissible, giving

\begin{equation}
\Psi_{n+1}
=
\frac{1}{2}
\left(
P_n
+
\sqrt{P_n^2+4\Psi_n}
\right).
\label{eq:psimap}
\end{equation}

Equation~(\ref{eq:psimap}) defines the radial update of the conservative Tokamap. Because it is derived directly from the Hamiltonian formulation without introducing any dissipative terms, the resulting transformation preserves phase-space area exactly, ensuring that invariant tori, magnetic islands, and chaotic regions arise solely from the intrinsic nonlinear dynamics of magnetic field-line transport.

The angular coordinate is obtained by discretizing Eq.~(\ref{eq:dtheta}) using the updated radial coordinate $\Psi_{n+1}$, yielding

\begin{equation}
T_{n+1}
=
T_n
+
\frac{1}{q(\Psi_{n+1})}
-
\frac{x_L}{4\pi^2}
\frac{\cos(2\pi T_n)}
{\left(1+\Psi_{n+1}\right)^2}.
\label{eq:Tmap}
\end{equation}

Because the poloidal angle is periodic, it is taken modulo unity after each iteration,

\begin{equation}
T_{n+1}
\leftarrow
T_{n+1}
\pmod 1.
\label{eq:mod}
\end{equation}

The magnetic equilibrium considered throughout this work is characterized by the safety-factor profile

\begin{equation}
q(\Psi)
=
x_q\left(1+\Psi^2\right)^2,
\label{eq:q}
\end{equation}
where the parameter $x_q$ controls both the magnitude and radial variation of the safety factor. This profile is monotonic and satisfies the basic requirements for describing a conventional tokamak equilibrium while remaining sufficiently simple to permit a systematic investigation of transport properties.

Combining Eqs.~(\ref{eq:psimap}) and (\ref{eq:Tmap}) yields the conservative Tokamap,

\begin{equation}
\left\{
\begin{array}{l}
\displaystyle
\Psi_{n+1}
=
\frac12
\left(
P_n+
\sqrt{P_n^2+4\Psi_n}
\right),
\\[0.5cm]
\displaystyle
T_{n+1}
=
T_n
+
\frac{1}{q(\Psi_{n+1})}
-
\frac{x_L}{4\pi^2}
\frac{\cos(2\pi T_n)}
{\left(1+\Psi_{n+1}\right)^2}
\pmod1,
\end{array}
\right.
\label{eq:Tokamap}
\end{equation}
with
\begin{equation}
P_n
=
\Psi_n
-
1
-
\frac{x_L}{2\pi}
\sin(2\pi T_n).
\label{eq:Pfinal}
\end{equation}

The mapping defined by Eq.~(\ref{eq:Tokamap}) constitutes an exact symplectic transformation. Consequently, the phase-space area is conserved under the iteration of the map, preserving the Hamiltonian nature of magnetic field-line dynamics. This property allows the Tokamap to faithfully reproduce the principal geometrical structures of the continuous system, including invariant tori, resonant island chains, cantori, and extended chaotic seas, without introducing artificial numerical diffusion.

The dynamics are governed by two independent control parameters. The parameter $x_L$ determines the amplitude of the magnetic perturbation and therefore controls the degree of resonance overlap. For small values of $x_L$, the phase space is dominated by invariant magnetic surfaces separated by isolated island chains. As $x_L$ increases, neighboring resonances progressively overlap, giving rise to stochastic layers and eventually to extended chaotic regions that facilitate radial transport.

The parameter $x_q$ specifies the safety-factor profile and therefore determines the magnetic shear. Variations in $x_q$ modify the local winding rate of magnetic field lines, changing the location of resonant surfaces and strongly influencing the organization of phase space, the size of regular islands, and the efficiency of chaotic transport. Throughout this work, these two parameters are varied independently to investigate the geometrical organization of phase space, the onset of chaos, scaling properties, and long-time transport in the conservative Tokamap.

\begin{figure}[t]
    \centering

    \begin{minipage}{0.49\columnwidth}
        \centering
        (a) \includegraphics[width=\linewidth]{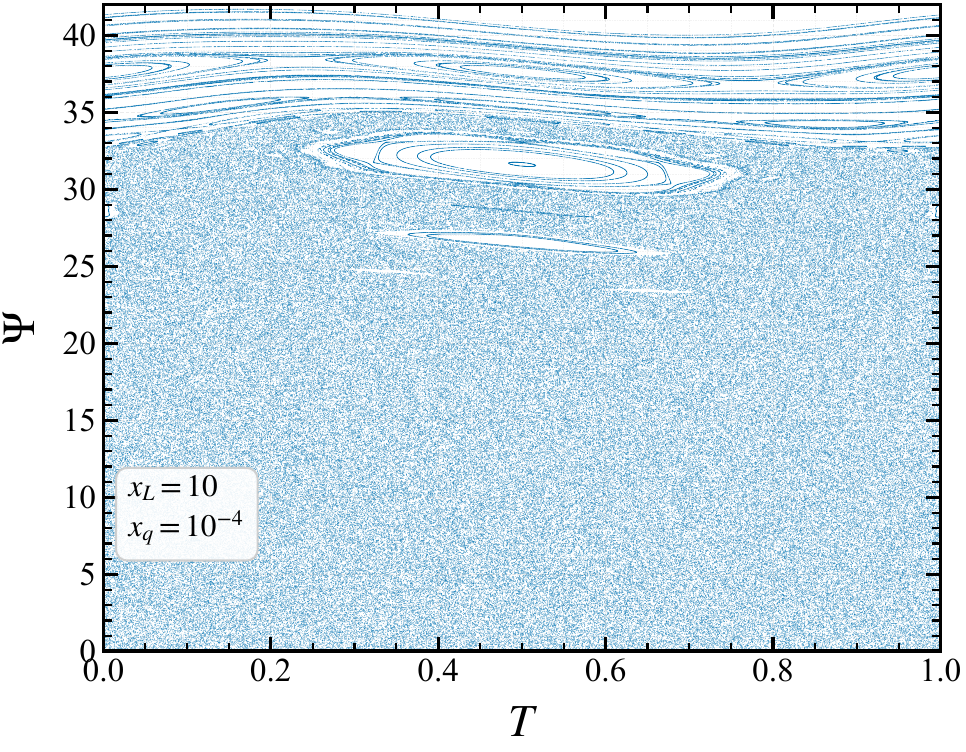}
    \end{minipage}
    \hfill
    \begin{minipage}{0.49\columnwidth}
        \centering
        (b) \includegraphics[width=\linewidth]{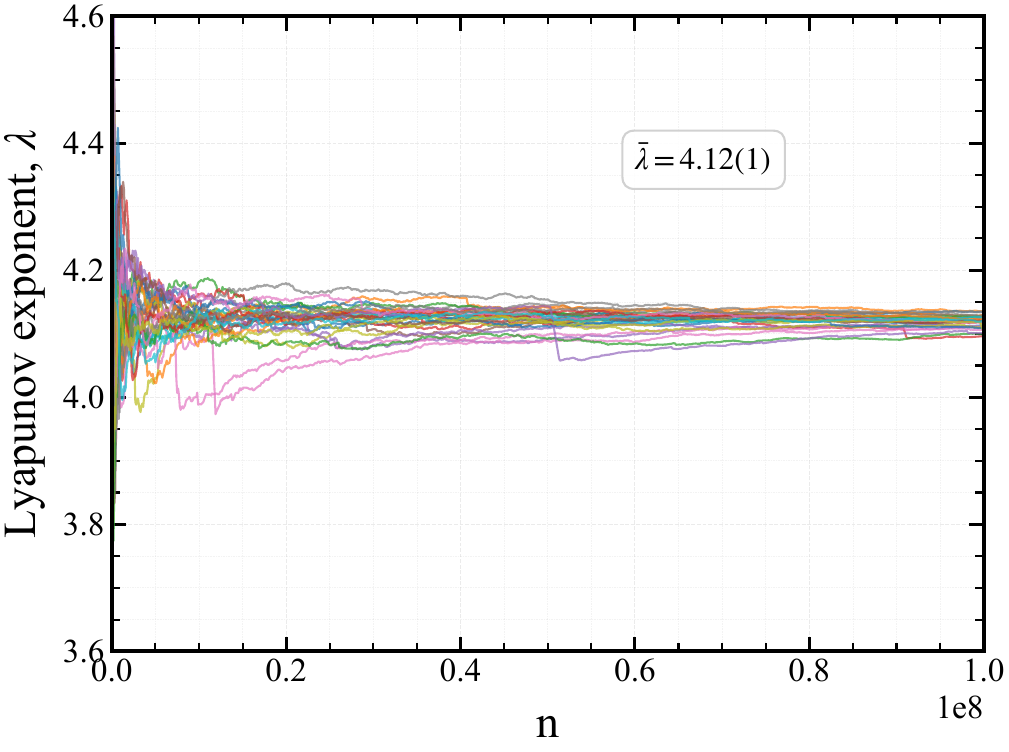}
    \end{minipage}

  \caption{ (a) Phase-space portrait showing invariant spanning curves, KAM islands, and a chaotic sea. The spanning curves act as transport barriers, the islands correspond to regular motion around stable periodic orbits, and the chaotic sea consists of irregular trajectories exhibiting stochastic transport. (b) Distribution of the largest Lyapunov exponent, yielding a mean value of $\bar{\lambda}=4.12 \pm 0.01$, consistent with chaotic behavior. In both images we considered the control parameters used are: $\delta=10^{-6}$ and $x_L=10$ .}
    \label{Fig2}
\end{figure}

Figure~\ref{Fig2}(a) presents a representative phase-space portrait of the conservative Tokamap, obtained by iterating Eqs.~\ref{eq:Tokamap} over a set of initial conditions. The coexistence of invariant spanning curves, magnetic island chains, and extended chaotic regions illustrates the mixed Hamiltonian character of the system. While invariant curves act as effective transport barriers, trajectories inside the chaotic sea exhibit irregular motion and can undergo long-range radial transport.

To quantify the degree of chaos, we compute the largest Lyapunov exponent (LLE), which measures the average exponential rate at which two initially nearby trajectories diverge in phase space~\cite{eckmann1985ergodic}. A positive largest Lyapunov exponent is the standard signature of chaotic dynamics, indicating sensitive dependence on initial conditions, whereas a vanishing exponent characterizes regular quasiperiodic motion confined to invariant tori.

The Lyapunov exponents are obtained from the linearized dynamics of the mapping. Let

\begin{equation}
\mathbf{X}_n
=
(\Psi_n,T_n)
\end{equation}
denote the phase-space vector after the $n$-th iteration of the Tokamap. The evolution of an infinitesimal perturbation,
$\delta\mathbf{X}_n$,
is governed by the tangent map,

\begin{equation}
\delta\mathbf{X}_{n+1}
=
J(\mathbf{X}_n)
\,
\delta\mathbf{X}_n,
\label{eq:tangent}
\end{equation}
where
\begin{equation}
J(\mathbf{X}_n)
=
\frac{\partial(\Psi_{n+1},T_{n+1})}
{\partial(\Psi_n,T_n)}
\label{eq:jacobian}
\end{equation}
is the Jacobian matrix of the Tokamap evaluated along the trajectory.

The Lyapunov spectrum is computed by repeatedly evolving the tangent vectors together with the orbit while periodically performing QR orthogonalization to prevent numerical overflow and preserve orthogonality of the perturbation vectors~\cite{benettin1980lyapunov}. The Lyapunov exponents are then given by

\begin{equation}
\lambda_j
=
\lim_{N\rightarrow\infty}
\frac{1}{N}
\sum_{i=1}^{N}
\ln
\left|
R_{jj}^{(i)}
\right|,
\qquad
j=1,2,
\label{eq:lyapunov}
\end{equation}
where $R_{jj}^{(i)}$ are the diagonal elements of the upper triangular matrix obtained during the QR decomposition at the $i$-th orthogonalization step.

Because the Tokamap is an exact symplectic mapping, the Jacobian satisfies $\det J=1$,implying that the Lyapunov exponents occur in conjugate pairs, $\lambda_1+\lambda_2=0.$ Consequently, the dynamics are completely characterized by the largest Lyapunov exponent, since the second exponent is simply its negative.

Figure~\ref{Fig2}(b) shows the convergence of the largest Lyapunov exponent for 30 initial conditions chosen within the chaotic sea. After a sufficiently large number of iterations, all trajectories converge to the same positive asymptotic value, $\bar{\lambda} = 4.12 \pm 0.01$, confirming that the selected region corresponds to fully developed chaotic dynamics. This positive Lyapunov exponent provides a quantitative characterization of the chaotic transport observed in the phase-space portrait and serves as the basis for the scaling and transport analyses presented in the following sections.

\begin{figure}[t]
    \centering

    \begin{minipage}{0.49\columnwidth}
        \centering
        (a)        \includegraphics[width=\linewidth]{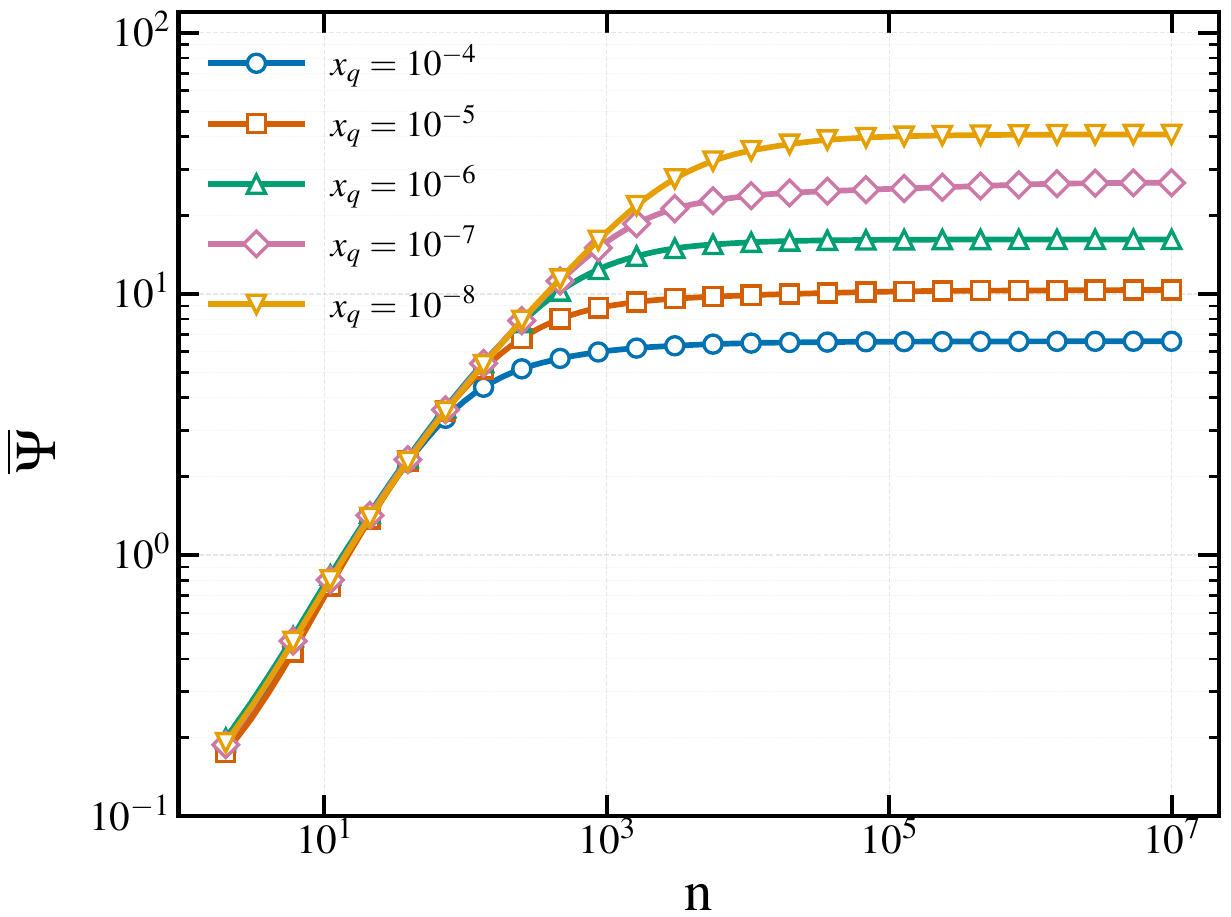}
    \end{minipage}
    \hfill
    \begin{minipage}{0.49\columnwidth}
        \centering
        (b)
        \includegraphics[width=\linewidth]{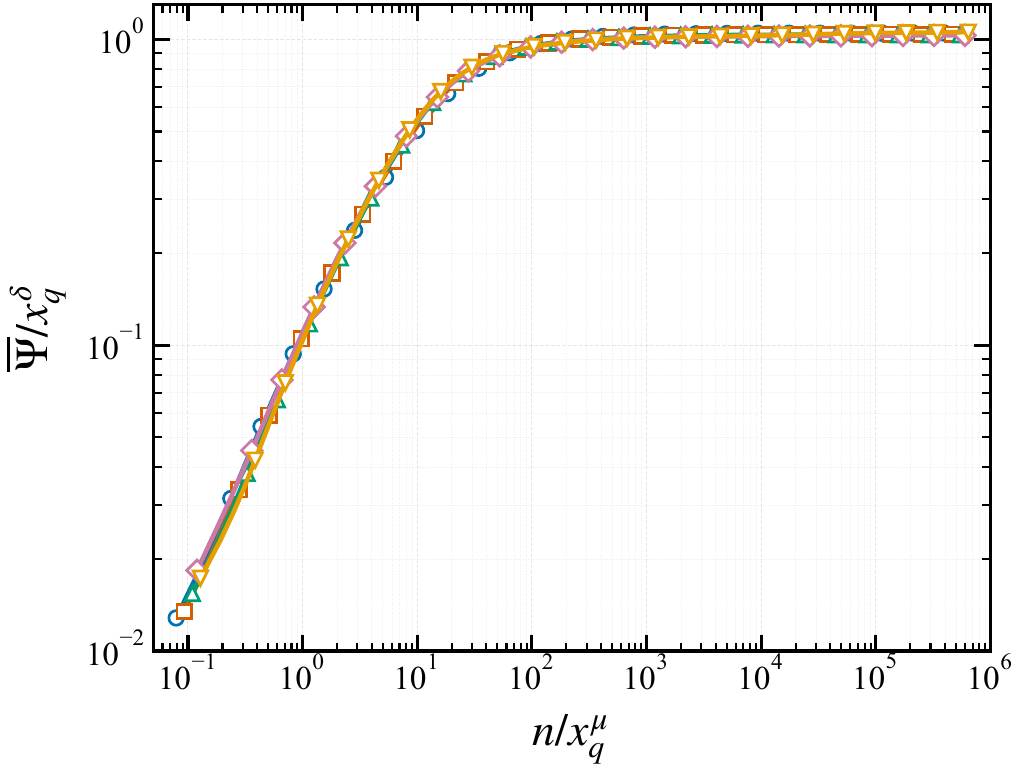}
    \end{minipage}

    \caption{(a) Different curves of $\overline{\Psi}$ for five different values of the control parameter $x_q$. (b) Collapse of the data onto a single universal curve after the appropriate scaling transformations.}
    \label{Phivs_n}
\end{figure}

\subsection{Ensemble-Averaged Transport Dynamics}

Chaotic transport in Hamiltonian systems is intrinsically statistical. Although trajectories initiated from nearby initial conditions obey the same deterministic equations of motion, they may explore distinct regions of phase space over finite times due to the sensitive dependence on initial conditions. Consequently, meaningful transport properties cannot be inferred from a single trajectory but instead require averaging over a sufficiently large ensemble of independent realizations. In this section, we characterize the global transport properties of the conservative Tokamap by analyzing the ensemble-averaged radial coordinate and its temporal evolution.

For each realization, the radial coordinate is first averaged over the entire trajectory. The resulting quantity is then averaged over all realizations, yielding

\begin{equation}
\overline{\Psi}
=
\frac{1}{Z}
\sum_{i=1}^{Z}
\left[
\frac{1}{N}
\sum_{n=1}^{N}
\Psi_i(n)
\right],
\label{eq:diversity}
\end{equation}
where $N$ denotes the total number of iterations, $Z$ is the number of independent initial conditions, and $\Psi_i(n)$ is the radial coordinate of the $i$-th trajectory after the $n$-th iteration.

Throughout this section, the perturbation parameter is fixed at $x_L=10$, corresponding to a fully developed chaotic regime in which large-scale radial transport is observed. Numerical results are obtained from an ensemble of $10^{4}$ independent trajectories. All trajectories are initialized at the same radial position, $\Psi_0 = 5\times10^{-5}$, while the initial poloidal angle is uniformly distributed over the interval $T_0\in[0,1).$

Figure~\ref{Phivs_n}(a) presents the evolution of the ensemble-averaged radial coordinate as a function of the number of iterations for different values of the magnetic-shear parameter $x_q$. Three distinct dynamical regimes are clearly observed. During the initial stage, the average radial displacement increases according to a power law, indicating rapid diffusion of trajectories through the chaotic sea. This regime is followed by a crossover region in which the transport gradually slows before approaching a statistically stationary state characterized by a constant saturation value.

The magnetic-shear parameter strongly influences the transport dynamics. Increasing $x_q$ shifts the crossover toward smaller iteration numbers and reduces the saturation level, indicating that stronger magnetic shear limits the radial exploration of chaotic trajectories. Conversely, smaller values of $x_q$ allow trajectories to diffuse over larger regions of phase space before reaching statistical equilibrium.

The transport dynamics presented in Fig.~\ref{Phivs_n}(a) exhibit the characteristic features of a system approaching a nonequilibrium stationary state through a scale-invariant process. The coexistence of a power-law growth regime, a well-defined crossover, and a stationary saturation regime suggests that the radial transport is governed by dynamic scaling, a behavior commonly observed in nonlinear systems displaying self-similar evolution~\cite{oliveira2009scaling,oliveira2015symmetry}. Motivated by these observations, we formulate three scaling hypotheses that describe the dependence of the transport dynamics on the magnetic-shear parameter.

The first hypothesis characterizes the early-time evolution. Before the crossover iteration, the ensemble-averaged radial coordinate grows as a power law,

\begin{equation}
\overline{\Psi}(n,x_q)
\propto
n^{\beta},
\qquad
n\ll n_x,
\label{eq:decay}
\end{equation}
where $\beta$ is the growth exponent. This exponent quantifies the rate at which chaotic trajectories spread through phase space during the initial transport regime.

The second hypothesis describes the long-time behavior. After sufficiently many iterations, the transport reaches a statistically stationary state in which the ensemble-averaged radial coordinate fluctuates around a constant saturation value,

\begin{equation}
\Psi_{\mathrm{sat}}
\propto
x_q^{\delta},
\qquad
n\gg n_x,
\label{eq:saturation}
\end{equation}
where $\delta$ is the saturation exponent. This exponent measures how the asymptotic extent of radial transport depends on the magnetic shear.

The third hypothesis concerns the characteristic iteration at which the dynamics change from the growth regime to the stationary regime. We assume that the crossover iteration obeys

\begin{equation}
n_x
\propto
x_q^{z},
\label{eq:crossover}
\end{equation}
where $z$ is the dynamic crossover exponent.

Together, these three hypotheses completely characterize the transport dynamics through three independent critical exponents. The exponent $\beta$ describes the initial rate of chaotic transport, $\delta$ determines the asymptotic transport level, and $z$ specifies the characteristic time scale required for the system to approach the stationary state. In the following subsection, we show that these exponents are not independent but are connected through a homogeneous scaling function, leading to a scaling law that provides a quantitative description of the transport dynamics.

The scaling hypotheses introduced above suggest that the transport dynamics are self-similar under suitable rescaling of the iteration number and the control parameter. This invariance can be expressed by assuming that the ensemble-averaged radial coordinate is a generalized homogeneous function,

\begin{equation}
\overline{\Psi}(n,x_q)
=
\ell\,
F\!\left(
\ell^{a}n,
\ell^{b}x_q
\right),
\label{eq:homogeneous}
\end{equation}
where $\ell$ is an arbitrary scaling factor, $F$ is a universal scaling function, and $a$ and $b$ are scaling exponents. Different choices of the scaling factor allow the asymptotic behavior of the transport dynamics to be recovered in the different dynamical regimes.

To describe the early-time dynamics, we choose $\ell=n^{-1/a}$, which transforms Eq.~(\ref{eq:homogeneous}) into

\begin{equation}
\overline{\Psi}(n,x_q)
=
n^{-1/a}
\Psi_1
\left(
1,
n^{-b/a}x_q
\right),
\label{eq:homogeneous_growth}
\end{equation}
where $\Psi_1$ is another universal scaling function. In the growth regime ($n\ll n_x$), the second argument of $\Psi_1$ varies slowly and may therefore be regarded as approximately constant. Consequently,

\begin{equation}
\overline{\Psi}
\propto
n^{-1/a}.
\end{equation}

Comparison with the first scaling hypothesis,
Eq.~(\ref{eq:decay}),
immediately yields

\begin{equation}
\beta
=
-\frac{1}{a}.
\label{eq:beta}
\end{equation}

To investigate the stationary regime, we instead choose $ \ell = x_q^{-1/b}$, which gives

\begin{equation}
\overline{\Psi}(n,x_q)
=
x_q^{-1/b}
\Psi_2
\left(
x_q^{-a/b}n,
1
\right),
\label{eq:homogeneous_sat}
\end{equation}
where $\Psi_2$ is again assumed to approach a constant for $n\gg n_x$.
The stationary transport therefore satisfies

\begin{equation}
\overline{\Psi}
\propto
x_q^{-1/b},
\end{equation}
which, when compared with Eq.~(\ref{eq:saturation}), leads to

\begin{equation}
\delta
=
-\frac{1}{b}.
\label{eq:delta}
\end{equation}

Finally, the crossover iteration is obtained by equating the two expressions for the scaling factor, $n^{-1/a}
= x_q^{-1/b}$, from which it follows that $n_x \propto x_q^{a/b}.$

\begin{figure}[t]
    \centering
    \begin{minipage}{0.49\columnwidth}
        \centering
        (a)  \includegraphics[width=\linewidth]{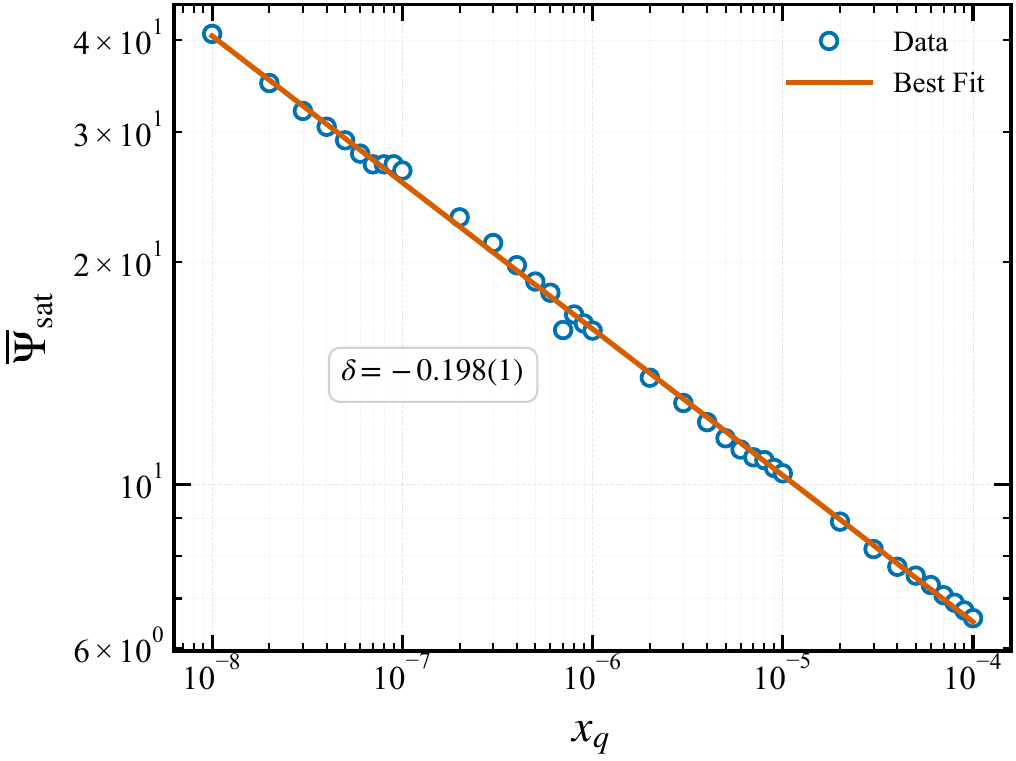}
    \end{minipage}
    \hfill
    \begin{minipage}{0.49\columnwidth}
        \centering
        (b)\includegraphics[width=\linewidth]{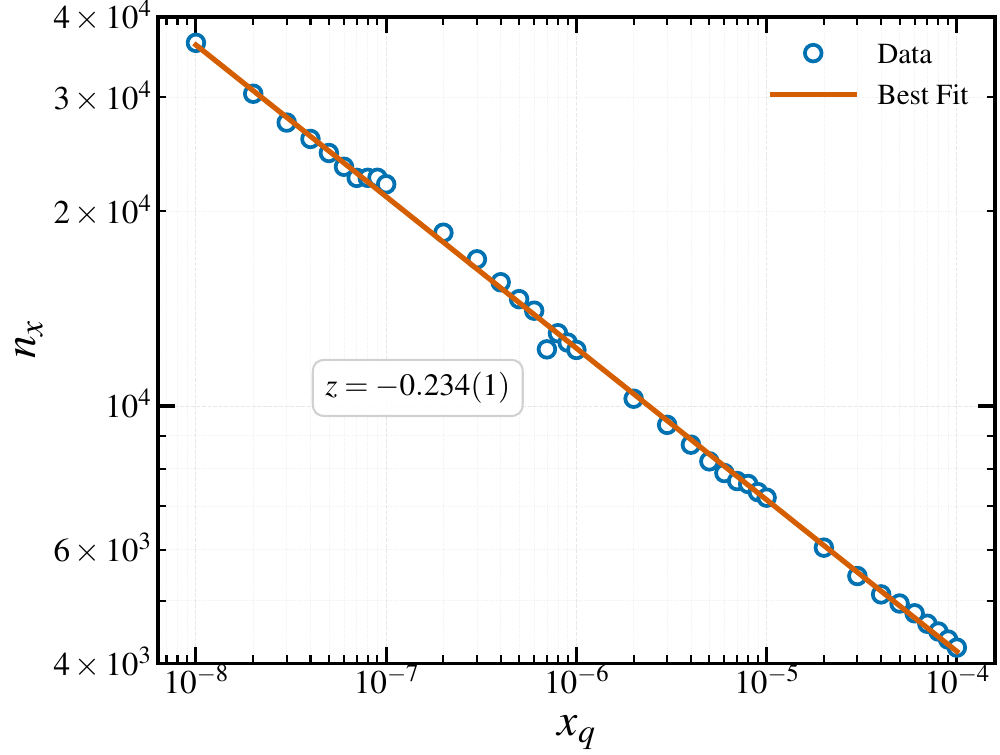}
    \end{minipage}

    \caption{(a) Behavior of the saturation value $\Psi_{\mathrm{sat}}$ as a function of $x_q$. (b) Behavior of the crossover iteration number $n_x$ as a function of $x_q$. Power-law fits yield the critical exponents $\delta=-0.198(1)$ in (a) and $z=-0.234(1)$ in (b).}
    \label{fig:expo}
\end{figure}

Using Eqs.~(\ref{eq:beta}) and (\ref{eq:delta}), the exponent ratio becomes

\begin{equation}
\frac{a}{b}
=
\frac{\delta}{\beta},
\end{equation}
and comparison with Eq.~(\ref{eq:crossover}) yields the scaling law

\begin{equation}
z
=
\frac{\delta}{\beta}.
\label{eq:scalinglaw}
\end{equation}

Equation~(\ref{eq:scalinglaw}) demonstrates that only two of the three critical exponents are independent. Once the growth and saturation exponents are determined, the crossover exponent is completely specified by the scaling theory. In the next subsection, these predictions are tested by independently determining the critical exponents from numerical simulations and verifying the resulting universal scaling collapse.

The scaling theory predicts that the transport dynamics are completely characterized by three critical exponents related through the scaling law given by Eq.~(\ref{eq:scalinglaw}). To validate this prediction, we determine each exponent independently from the numerical simulations and compare the resulting values with the theoretical expectation.

The growth exponent $\beta$ is obtained from a least-squares fit of the power-law regime shown in Fig.~\ref{Phivs_n}(a). Averaging over all realizations yields $\beta=0.847(9),$ where the uncertainty corresponds to the standard error of the fit.

The saturation exponent is determined from the dependence of the stationary value,
$\Psi_{\mathrm{sat}}$,
on the magnetic-shear parameter. As shown in Fig.~\ref{fig:expo}(a), the numerical data are accurately described by the power law

\begin{equation}
\Psi_{\mathrm{sat}}
\propto
x_q^{\delta},
\end{equation}
from which we obtain $\delta=-0.198(1)$ as shown in Fig. \ref{fig:expo} (a).

Similarly, the crossover iteration is extracted from the intersection between the asymptotic power-law growth and the saturation plateau. Figure~\ref{fig:expo}(b) shows that the crossover obeys

\begin{equation}
n_x
\propto
x_q^{z},
\end{equation}
yielding $z=-0.234(2)$ as shown in Fig. \ref{fig:expo} (b).

An independent estimate of the crossover exponent can be obtained directly from the scaling law,

\begin{equation}
z
=
\frac{\delta}{\beta},
\end{equation}
which, using the independently measured values of $\beta$ and $\delta$, gives $z=-0.233(3).$

The excellent agreement between the directly measured value and the prediction of the scaling theory demonstrates the internal consistency of the proposed scaling framework and confirms that the transport dynamics are governed by only two independent critical exponents.

A more stringent test of the scaling hypotheses is provided by the collapse of the transport curves onto a single universal function. According to the homogeneous scaling relation, the variables are rescaled as

\begin{equation}
n
\rightarrow
\frac{n}{x_q^{\,z}},
\qquad
\overline{\Psi}
\rightarrow
\frac{\overline{\Psi}}
{x_q^{\,\delta}}.
\end{equation}

The resulting rescaled curves are shown in Fig.~\ref{Phivs_n}(b). Despite spanning different values of the magnetic-shear parameter, all data collapse onto a single master curve over the entire dynamical range, including the growth, crossover, and saturation regimes. This universal collapse provides strong evidence that the transport dynamics obey dynamic scaling and confirms the validity of the proposed scaling theory.

The existence of a universal scaling function further indicates that the long-time transport properties of the conservative Tokamap are largely independent of the specific value of the magnetic-shear parameter. Instead, the dynamics are governed by the critical exponents and the associated scaling function, which completely characterize the evolution from the initial diffusion regime to the statistically stationary state.

\begin{figure}[t]
    \centering
    \begin{minipage}{0.49\columnwidth}
        \centering
        (a)  \includegraphics[width=\linewidth]{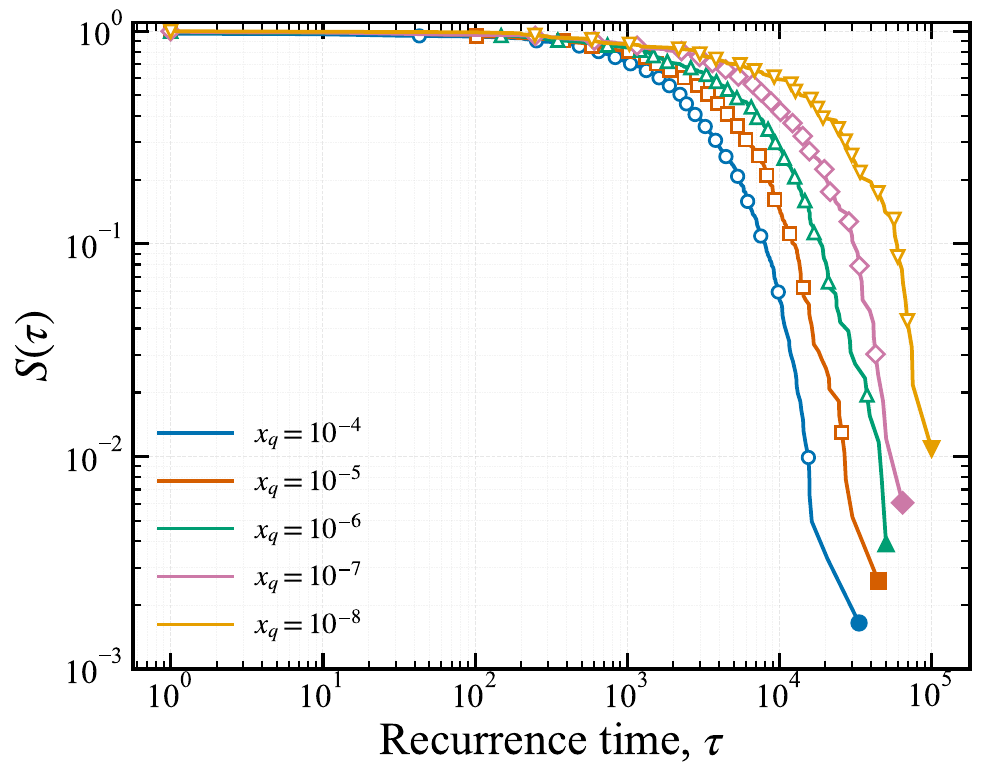}
    \end{minipage}
    \hfill
    \begin{minipage}{0.49\columnwidth}
        \centering
        (b) \includegraphics[width=\linewidth]{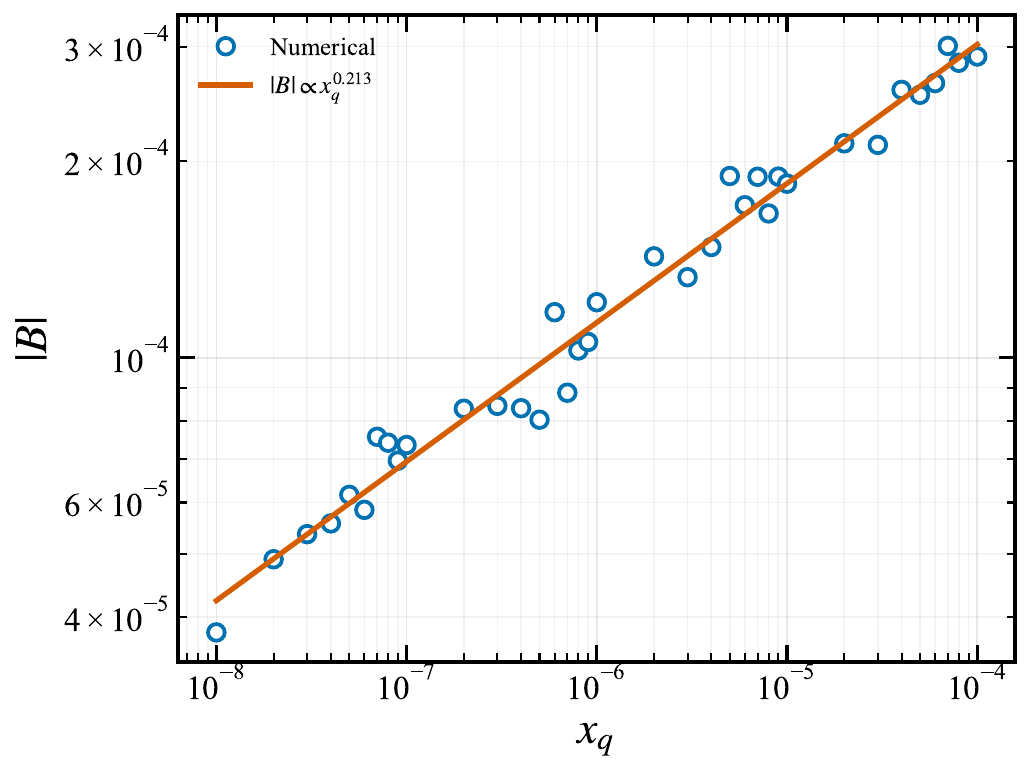}
    \end{minipage}

    \caption{Recurrence statistics for the conservative Tokamap. (a) Survival probability $S(\tau)$ for different values of the perturbation parameter $x_q$, computed using a recurrence box centered at $(T,\Psi)=(0.5,5.0)$ with $\Delta T=0.01$ and $\Delta\Psi=0.10$. The systematic shift of the curves toward longer recurrence times as $x_q$ decreases indicates progressively slower chaotic transport and stronger trapping near invariant structures. (b) Absolute value of the exponential decay coefficient $|B|$, obtained by fitting the initial portion of the survival probability ($S(\tau)>0.08$) to $S(\tau)=P_0e^{B\tau}$. The solid line shows the power-law fit, $|B|\propto x_q^{0.213}$, demonstrating that the characteristic transport time increases continuously as the perturbation weakens.}

    \label{Surv}
\end{figure}

\subsection{Intermittent Transport and Recurrence Statistics}

The results presented in the previous sections establish the geometrical and dynamical properties of the conservative Tokamap. Phase-space portraits revealed the coexistence of regular and chaotic regions, the Lyapunov exponents quantified the local instability of chaotic trajectories, and the scaling analysis characterized the evolution of radial transport. Together, these diagnostics provide a detailed description of the short- and intermediate-time dynamics. However, they do not fully characterize the mechanisms governing transport over asymptotically long time scales.

In Hamiltonian systems with mixed phase space \cite{lichtenberg1992,meiss1992,mackay1984}, long-time transport is determined not only by the existence of chaotic trajectories but also by the complex network of invariant structures embedded within the chaotic sea. Kolmogorov-Arnold-Moser (KAM) tori, resonance island chains, and cantori \cite{mackay1984,meiss1986} do not completely prevent transport but instead act as partial barriers that repeatedly interrupt the motion of chaotic trajectories. As a result, trajectories alternate between relatively rapid excursions through the chaotic sea and long episodes during which they remain temporarily trapped near regular structures before escaping and continuing their motion. This phenomenon, commonly referred to as \emph{stickiness} \cite{karney1983,zaslavsky2002,chirikov1999}, is one of the defining characteristics of transport in Hamiltonian systems and is responsible for the anomalously long exploration times observed in mixed phase spaces.

Because stickiness is inherently a long-time phenomenon, its effects cannot be inferred solely from local quantities such as Lyapunov exponents. Two trajectories with nearly identical local instability may exhibit remarkably different transport properties depending on the amount of time they spend trapped near invariant structures. Consequently, a statistical characterization of recurrence events is required to quantify the efficiency of transport throughout the chaotic sea.

Poincaré recurrence statistics \cite{altmann2013,chirikov1999} provide a natural framework for this analysis. Rather than measuring the instantaneous divergence of nearby trajectories, recurrence statistics quantify the time required for a chaotic trajectory to return to a prescribed region of phase space after exploring the surrounding chaotic component. These return times directly probe the influence of sticky motion and therefore provide a sensitive measure of the transport efficiency. Short recurrence times indicate rapid exploration of phase space, whereas long recurrence times reveal prolonged trapping near invariant structures and, consequently, reduced transport.

The objective of this section is to establish a direct quantitative connection between the geometrical organization of the phase space and the corresponding transport dynamics. By analyzing the statistics of Poincaré recurrences over a broad range of control parameters, we demonstrate how the progressive reinforcement of invariant structures continuously suppresses chaotic transport, leading to increasingly intermittent dynamics and systematically longer transport times.

To investigate the transport properties of the conservative Tokamap, we analyze the statistics of Poincaré recurrences generated by long chaotic trajectories. For each value of the control parameter $x_q$, a single orbit was iterated after discarding an initial transient of $10^{3}$ iterations to eliminate the influence of the initial conditions and ensure that the dynamics had reached the asymptotic regime.

A fixed recurrence region was selected inside the chaotic component of phase space, centered at
$(T_{\mathrm{box}},\Psi_{\mathrm{box}}) = (0.5,5.0),$
with half-widths $\Delta T=0.01$, and  $\Delta\Psi=0.10$.

A recurrence event was recorded whenever the trajectory satisfied

\begin{equation}
|T_n-T_{\mathrm{box}}|
\le
\Delta T,
\qquad
|\Psi_n-\Psi_{\mathrm{box}}|
\le
\Delta\Psi.
\label{eq:box}
\end{equation}

The same recurrence region was used for every simulation to ensure that changes in the recurrence statistics originated exclusively from variations in the control parameter rather than from the location of the observation window.

The recurrence time was defined as the interval between two consecutive visits to the recurrence region,

\begin{equation}
\tau_i
=
n_{i+1}-n_i,
\label{eq:tau}
\end{equation}
where $n_i$ denotes the iteration corresponding to the $i$th return. The sequence of recurrence times therefore provides a direct measure of the efficiency with which chaotic trajectories revisit a given region of phase space after exploring the surrounding chaotic sea.

Rather than analyzing the recurrence-time distribution itself, we consider its complementary cumulative distribution, commonly referred to as the survival probability,

\begin{equation}
S(\tau)
=
P(T>\tau),
\label{eq:survival}
\end{equation}
which gives the probability that a recurrence time exceeds a prescribed value $\tau$. Compared with conventional recurrence-time histograms, the survival probability \cite{altmann2013} is considerably less sensitive to statistical fluctuations, particularly in the long-time regime where recurrence events become increasingly rare. Consequently, it provides a more robust characterization of transport over extended time scales.

From a dynamical perspective, the long-time tail of the survival probability contains direct information about sticky motion \cite{karney1983,zaslavsky2002}. Rapidly decaying survival curves indicate efficient transport through the chaotic sea, whereas slowly decaying tails reveal that trajectories spend long intervals trapped near invariant structures before returning to the recurrence region. The survival probability therefore provides a quantitative bridge between the geometrical organization of the phase space and the corresponding transport dynamics, making it an ideal observable for investigating long-time transport in the conservative Tokamap.

Figure~\ref{Surv}(a) presents the survival probability for several values of the magnetic-shear parameter $x_q$. A clear and systematic dependence on the control parameter is observed throughout the entire range of recurrence times. As $x_q$ decreases from $10^{-4}$ to $10^{-8}$, the survival curves are progressively displaced toward larger recurrence times, indicating that chaotic trajectories require increasingly longer intervals to revisit the same region of phase space. This monotonic displacement immediately demonstrates that the characteristic transport time increases continuously as the magnetic shear is reduced.

The effect of the control parameter is particularly evident in the decay of the survival probability. For the largest value investigated, $x_q=10^{-4}$, the survival probability decreases rapidly, indicating that trajectories efficiently explore the chaotic sea and return frequently to the recurrence region. As $x_q$ decreases, the decay becomes progressively slower over the entire time interval. In particular, the curve corresponding to $x_q=10^{-8}$ exhibits the slowest decay, with a substantially enhanced probability of long recurrence events. The preservation of the ordering of all curves throughout the investigated interval demonstrates that the slowing of transport is continuous rather than the result of an abrupt dynamical transition.

This behavior has a direct dynamical interpretation. Reducing $x_q$ modifies the magnetic-shear profile, thereby increasing the influence of invariant structures embedded within the chaotic sea. Although chaotic trajectories remain globally connected, they experience progressively longer trapping episodes near KAM islands, resonance chains, and cantori before escaping and continuing their motion. These intermittent trapping events, characteristic of sticky dynamics, substantially increase the recurrence times and reduce the overall efficiency of transport.

The survival probability therefore provides direct quantitative evidence of the increasing importance of stickiness as the magnetic shear is reduced. While phase-space portraits reveal the geometrical organization of invariant structures and Lyapunov exponents quantify the local instability of trajectories, recurrence statistics capture the cumulative dynamical consequences of these structures over long time scales. The systematic displacement of the survival curves thus establishes a direct connection between the geometry of the mixed phase space and the progressive suppression of chaotic transport.

To quantify the evolution of the transport rate, the initial portion of each survival curve, corresponding to $S(\tau)>0.08$, was fitted by the exponential function

\begin{equation}
S(\tau)=P_{0}e^{B\tau},
\label{eq:expfit}
\end{equation}
where $P_{0}$ and $B$ are fitting parameters. Restricting the fit to this interval minimizes the influence of statistical fluctuations associated with the sparsely populated long-time tail while accurately describing the dominant transport regime observed in the simulations.

The fitted decay coefficients are shown in Fig.~\ref{Surv}(b). Over nearly five decades in the magnetic-shear parameter, the numerical results follow the scaling law

\begin{equation}
|B|\propto x_q^{0.213},
\label{eq:Bscaling}
\end{equation}
indicating that the recurrence dynamics are governed by a well-defined algebraic dependence on the control parameter. Since the characteristic transport time associated with the exponential regime is inversely proportional to the decay rate,

\begin{equation}
\tau_c=\frac{1}{|B|},
\end{equation}
the previous relation immediately yields

\begin{equation}
\tau_c\propto x_q^{-0.213}.
\label{eq:tauscaling}
\end{equation}

This result demonstrates that the characteristic transport time increases continuously as the magnetic shear decreases. Although trajectories remain globally chaotic throughout the investigated parameter interval, they require progressively longer times to explore the accessible phase space because sticky episodes become increasingly dominant. Consequently, the efficiency of chaotic transport is controlled not only by local exponential instability but also by the global organization of the phase space.

The recurrence analysis completes the physical picture developed throughout this work. Phase-space portraits revealed the evolution of the mixed phase space, Lyapunov exponents quantified the local instability of chaotic trajectories, the scaling analysis characterized the growth of radial transport, and the recurrence statistics extended these results to asymptotically long times by directly measuring the transport dynamics. Together, these complementary diagnostics consistently demonstrate that reducing the magnetic shear progressively suppresses transport through the reinforcement of invariant structures embedded within the chaotic sea.

More importantly, the present results highlight a fundamental aspect of transport in Hamiltonian systems: chaos and transport are not equivalent concepts. While positive Lyapunov exponents guarantee exponential separation of nearby trajectories, they do not necessarily imply efficient exploration of phase space. Instead, long-time transport is governed by the repeated interaction of chaotic trajectories with KAM islands, resonance chains, and cantori, which act as partial transport barriers \cite{mackay1984,meiss1986} and generate intermittent dynamics through stickiness. The algebraic scaling of the characteristic transport time provides direct quantitative evidence of this mechanism, establishing a clear connection between the microscopic phase-space organization and the macroscopic transport properties of the system.

Beyond the conservative Tokamap, these findings emphasize the broader utility of recurrence statistics as a quantitative framework for investigating transport in Hamiltonian systems with mixed phase space. By linking invariant geometrical structures to measurable transport times, the present approach provides new insight into the mechanisms regulating magnetic-field-line transport \cite{balescu1998tokamap,viana2023hamiltonian,balescu2005,wootton1990fluctuations} in toroidal plasmas and offers a general methodology for studying intermittent transport in conservative nonlinear systems. Future work will extend this analysis to dissipative Tokamaps and magnetic configurations with reversed magnetic shear, where attractor dynamics and crisis-induced intermittency are expected to produce qualitatively different transport regimes.

\section{Conclusions}

In this work, we presented a comprehensive investigation of magnetic field-line transport in the conservative Tokamap by combining geometrical, dynamical, statistical, and scaling analyses within a unified Hamiltonian framework. Starting from the Hamiltonian formulation of magnetic field-line dynamics, we derived the conservative Tokamap as an exact symplectic mapping and used it to investigate the interplay between magnetic shear, chaotic dynamics, and long-time transport.

The phase-space analysis revealed the characteristic mixed structure of the conservative Tokamap, consisting of invariant spanning curves, magnetic island chains, cantori, and extended chaotic regions. The largest Lyapunov exponent provided a quantitative characterization of local dynamical instability and confirmed the existence of fully developed chaotic motion over a broad range of control parameters. However, our results demonstrate that local instability alone is insufficient to characterize transport in Hamiltonian systems.

The ensemble-averaged radial transport exhibits a well-defined scaling behavior characterized by three critical exponents describing the growth, saturation, and crossover regimes. These exponents satisfy a scaling relation derived from a generalized homogeneous scaling hypothesis, and the excellent collapse of the transport curves onto a universal scaling function demonstrates that the long-time dynamics are governed by scale-invariant transport laws. This result establishes dynamic scaling as an effective framework for describing radial transport in the conservative Tokamap.

To characterize transport over asymptotically long time scales, we analyzed Poincar\'e recurrence statistics. The systematic displacement of the survival probabilities toward longer recurrence times, together with the algebraic scaling of the characteristic transport time, demonstrates that reducing the magnetic shear progressively suppresses transport by increasing the influence of invariant structures embedded within the chaotic sea. Although the dynamics remain globally chaotic, trajectories spend increasingly long intervals trapped near KAM islands, resonance chains, and cantori before continuing their exploration of phase space. These results provide direct quantitative evidence that sticky motion is the dominant mechanism governing long-time transport in the conservative Tokamap.

More broadly, the present study highlights a fundamental property of Hamiltonian transport: chaoticity and transport efficiency are distinct concepts. Positive Lyapunov exponents quantify the local exponential divergence of nearby trajectories, whereas global transport is ultimately controlled by the geometrical organization of phase space through partial transport barriers and intermittent dynamics. By combining phase-space geometry, Lyapunov analysis, dynamic scaling, and recurrence statistics within a single framework, this work establishes a direct connection between microscopic phase-space structures and macroscopic transport properties.

Because the conservative Tokamap preserves the Hamiltonian structure of magnetic field-line dynamics while remaining computationally efficient, it provides an ideal framework for investigating nonlinear transport in magnetically confined plasmas. The methodology developed here is readily applicable to more realistic magnetic configurations, including bounded and symmetric Tokamaps, reversed magnetic shear, ergodic divertors, and magnetic perturbations generated by resonant magnetic perturbation coils. Future work will extend the present approach to dissipative Tokamaps and non-twist magnetic configurations, where attractors, crises, and additional intermittent transport mechanisms are expected to produce qualitatively different transport regimes.


\begin{thebibliography}{10}

\bibitem{wesson2011}
Wesson J and Campbell D~J 2011 {\em Tokamaks\/} vol 149 (Oxford university
  press)

\bibitem{ongena2016magnetic}
Ongena J, Koch R, Wolf R and Zohm H 2016 {\em Nature Physics\/} {\bf 12}
  398--410

\bibitem{stacey2010fusion}
Stacey W~M 2010 {\em Fusion: An introduction to the physics and technology of
  magnetic confinement fusion\/} (John Wiley \& Sons)

\bibitem{miyamoto2005plasma}
Miyamoto K 2005 {\em Plasma physics and controlled nuclear fusion\/} (Springer)

\bibitem{white2014}
White R~B 2013 {\em Theory Of Toroidally Confined Plasmas, The\/} (World
  Scientific Publishing Company)

\bibitem{hazeltine2003}
Hazeltine R~D and Meiss J~D 2003 {\em Plasma confinement\/} (Courier
  Corporation)

\bibitem{freidberg2014}
Freidberg J~P 2014 {\em ideal MHD\/} (Cambridge University Press)

\bibitem{freidberg2008plasma}
Freidberg J~P 2008 {\em Plasma physics and fusion energy\/} (Cambridge
  university press)

\bibitem{rechester1978}
Rechester A and Rosenbluth M 2020 Electron heat transport in a tokamak with
  destroyed magnetic surfaces {\em Hamiltonian Dynamical Systems\/} (CRC Press)
  pp 684--7

\bibitem{abdullaev2006}
Abdullaev S~S 2006 {\em Construction of Mappings for Hamiltonian Systems and
  Their Applications\/} (Berlin: Springer)

\bibitem{evans2006edge}
Evans T~E, Moyer R~A, Burrell K~H, Fenstermacher M~E, Joseph I, Leonard A~W,
  Osborne T~H, Porter G~D, Schaffer M~J, Snyder P~B {\em et~al.\/} 2006 {\em
  nature physics\/} {\bf 2} 419--23

\bibitem{evans2006physics}
Evans T, Burrell K, Fenstermacher M, Moyer R, Osborne T, Schaffer M, West W,
  Yan L, Boedo J, Doyle E {\em et~al.\/} 2006 {\em Physics of plasmas\/} {\bf
  13}

\bibitem{loarte2007chapter}
Loarte A, Lipschultz B, Kukushkin A, Matthews G, Stangeby P, Asakura N,
  Counsell G, Federici G, Kallenbach A, Krieger K {\em et~al.\/} 2007 {\em
  Nuclear Fusion\/} {\bf 47} S203--63

\bibitem{kolmogorov1954}
Kolmogorov A~N 1954 On conservation of conditionally periodic motions for a
  small change in hamilton's function {\em Dokl. akad. nauk Sssr\/} vol~98 pp
  527--30

\bibitem{arnold1963}
Vladimir I 1963 {\em Russian Mathematical Surveys\/} {\bf 18} 85--191

\bibitem{moser1962}
M{\"o}ser J 1962 {\em Nachr. Akad. Wiss. G{\"o}ttingen, II\/}  1--20

\bibitem{chirikov1979}
Chirikov B~V 1979 {\em Physics reports\/} {\bf 52} 263--379

\bibitem{lichtenberg1992}
Lichtenberg A~J and Lieberman M~A 2013 {\em Regular and chaotic dynamics\/}
  vol~38 (Springer Science \& Business Media)

\bibitem{meiss1992}
Meiss J 1992 {\em Reviews of Modern Physics\/} {\bf 64} 795

\bibitem{leonel2016thermodynamics}
Leonel E~D, Galia M~V~C, Barreiro L~A and Oliveira D~F 2016 {\em Physical
  Review E\/} {\bf 94} 062211

\bibitem{greene1979}
Greene J~M 1979 {\em Journal of Mathematical Physics\/} {\bf 20} 1183--201

\bibitem{mackay1984}
MacKay R, Meiss J and Percival I 1984 {\em Physica D: Nonlinear Phenomena\/}
  {\bf 13} 55--81

\bibitem{zaslavsky2002}
Zaslavsky G~M 2002 {\em Physics reports\/} {\bf 371} 461--580

\bibitem{zaslavsky2007}
Zaslavsky G~M 2007 {\em The physics of chaos in Hamiltonian systems\/} (world
  scientific)

\bibitem{altmann2005stickiness}
Altmann E~G, Motter A~E and Kantz H 2005 {\em Chaos: An Interdisciplinary
  Journal of Nonlinear Science\/} {\bf 15}

\bibitem{boozer1983}
Boozer A~H 1979 Guiding center drift equations Tech. rep. Princeton Univ., NJ
  (USA). Plasma Physics Lab.

\bibitem{cary1983}
Cary J~R and Littlejohn R~G 1983 {\em Annals of Physics\/} {\bf 151} 1--34

\bibitem{constantinescu2005non}
Constantinescu D 2005 {\em Romanian Journal of Physics\/} {\bf 50} 325

\bibitem{hudson2012computation}
Hudson S, Dewar R, Dennis G, Hole M, McGann M, Von~Nessi G and Lazerson S 2012
  {\em Physics of Plasmas\/} {\bf 19}

\bibitem{abdullaev2006construction}
Abdullaev S~S 2006 {\em Construction of mappings for Hamiltonian systems and
  their applications\/} (Springer)

\bibitem{ullmann2000symplectic}
Ullmann K and Caldas I~L 2000 {\em Chaos, Solitons \& Fractals\/} {\bf 11}
  2129--40

\bibitem{morrison2000magnetic}
Morrison P 2000 {\em Physics of Plasmas\/} {\bf 7} 2279--89

\bibitem{abdullaev2014magnetic}
Abdullaev S {\em et~al.\/} 2014 {\em Magnetic stochasticity in magnetically
  confined fusion plasmas\/} vol~78 (Springer)

\bibitem{balescu1998tokamap}
Balescu R, Vlad M and Spineanu F 1998 {\em Physical Review E\/} {\bf 58} 951

\bibitem{eberhard2005symmetric}
Eberhard M 2005 {\em Physical Review E—Statistical, Nonlinear, and Soft
  Matter Physics\/} {\bf 71} 026411

\bibitem{bartoloni2016shearless}
Bartoloni B, Schelin A and Caldas I~L 2016 {\em Physics Letters A\/} {\bf 380}
  2416--21

\bibitem{balescu2003}
Balescu R 2005 {\em Aspects of anomalous transport in plasmas\/} (CRC Press)

\bibitem{wingen2005stochastic}
Wingen A, Spatschek K and Abdullaev S 2005 {\em Contributions to plasma
  physics\/} {\bf 45} 500--13

\bibitem{misguich2001dynamics}
Misguich J 2001 {\em Physics of Plasmas\/} {\bf 8} 2132--8

\bibitem{balescu1998revtokamap}
Balescu R 1998 {\em Physical Review E\/} {\bf 58} 3781

\bibitem{shinohara1998indicators}
Shinohara S and Aizawa Y 1998 {\em Progress of theoretical physics\/} {\bf 100}
  219--33

\bibitem{eckmann1985ergodic}
Eckmann J~P and Ruelle D 1985 {\em Reviews of modern physics\/} {\bf 57} 617

\bibitem{benettin1980lyapunov}
Benettin G, Galgani L, Giorgilli A and Strelcyn J~M 1980 {\em Meccanica\/} {\bf
  15} 9--20

\bibitem{oliveira2009scaling}
Oliveira D~F, Bizao R~A and Leonel E~D 2009 {\em Mathematical Problems in
  engineering\/} {\bf 2009} 213857

\bibitem{oliveira2015symmetry}
Oliveira D~F, Silva M~R and Leonel E~D 2015 {\em Physica A: Statistical
  Mechanics and its Applications\/} {\bf 436} 909--15

\bibitem{meiss1986}
Meiss J~D and Ott E 1986 {\em Physica D: Nonlinear Phenomena\/} {\bf 20}
  387--402

\bibitem{karney1983}
Karney C~F 1983 {\em Physica D: Nonlinear Phenomena\/} {\bf 8} 360--80

\bibitem{chirikov1999}
Chirikov B~V and Shepelyansky D~L 1984 {\em Physica D: Nonlinear Phenomena\/}
  {\bf 13} 395--400

\bibitem{altmann2013}
Altmann E~G and Kantz H 2005 {\em Physical Review E—Statistical, Nonlinear,
  and Soft Matter Physics\/} {\bf 71} 056106

\bibitem{viana2023hamiltonian}
Viana R~L, Mugnaine M and Caldas I~L 2023 {\em Physics of Plasmas\/} {\bf 30}

\bibitem{balescu2005}
Balescu R 2005 {\em Aspects of anomalous transport in plasmas\/} (CRC Press)

\bibitem{wootton1990fluctuations}
Wootton A, Carreras B, Matsumoto H, McGuire K, Peebles W, Ritz C~P, Terry P and
  Zweben S 1990 {\em Physics of Fluids B: Plasma Physics\/} {\bf 2} 2879--903

\end{thebibliography}

\providecommand{\newblock}{}
\providecommand{\url}[1]{{\tt #1}}
\providecommand{\urlprefix}{}
\providecommand{\href}[2]{#2}

\end{document}